\documentstyle[preprint,prl,aps,epsf]{revtex}

\newcommand{\up}{\uparrow}
\newcommand{\down}{\downarrow}

\begin{document}
\draft
\title{Comparison of Calculations for the Hubbard model obtained with
Quantum-Monte-Carlo, exact and stochastic Diagonalization}
\author{Thomas Husslein, Werner Fettes and Ingo Morgenstern}
\address{Universit\"at Regensburg, Fakult\"at Physik, 93040 Regensburg\\
 e-mail: thomas.husslein@physik.uni-regensburg.de; werner.fettes@physik.uni-regensburg.de}
\date{\today}

\maketitle

\begin{abstract}
In this paper we compare numerical results for the ground state
of the Hubbard model obtained by Quantum-Monte-Carlo simulations with 
results from 
exact and stochastic diagonalizations. We find good agreement for the ground state energy
and
superconducting correlations for both, the repulsive
and attractive Hubbard model. Special emphasis lies on the
superconducting correlations in the repulsive Hubbard model, 
where the small magnitude of the
values obtained by Monte-Carlo simulations gives rise to the
question, whether these results might be caused by fluctuations
or systematic errors of the method. Although we notice that the Quantum-Monte-Carlo
method has convergence problems for large interactions, coinciding with a minus sign
problem, we confirm the results of the diagonalization techniques for small and
moderate interaction strengths. 
Additionally we investigate the numerical stability and the convergence 
of the Quantum-Monte-Carlo method in the attractive case, to study the 
influence of the minus sign problem on convergence.
Also here in the absence of a minus sign problem we encounter convergence
problems for strong interactions.
\end{abstract}

\narrowtext
\section{Introduction}

The Quantum-Monte-Carlo-Algorithms (QMC) are a well established
tool in theoretical many particle physics \cite{Suz76} \cite{LIN92}. 
They have been successfully
used to investigate the spin and charge degrees of freedom for
the attractive and repulsive Hubbard model \cite{Hir84}. 
There are several kinds of QMC algorithms. We focus here on the Projector 
Quantum-Monte-Carlo-Algorithm (PQMC) \cite{KOO82}.
In the case of the repulsive Hubbard model (\,RHM\,) the sign problem severely
restricts the parameter space one can calculate. 
One goal of this paper is to show, what parameters of the RHM can 
be calculated using state of the art  QMC-methods 
and which of the observables still give
meaningful results when a sign problem occurs. 
Concerning the minus sign free attractive Hubbard model (\,AHM\,) 
we also want to distinguish numerical instabilities and convergence 
problems from well-"behaved" runs.

Inspired by results of de Raedt et al.\ \cite{Rae84} for
the Heisenberg model 
we additionally investigate whether the choice
of the elementary move in the Monte-Carlo procedure will have
influence on the results. 

We show that exact diagonalization (ED)~\cite{CUL85}, the stochastic
diagonalization(SD)~\cite{RAE92} and PQMC~\cite{KOO82} give consistent results 
for a broad range of the simulated parameters.

In our paper we investigated a modified version of the Hubbard model
the $t$-$t_p$-Hubbard model that is defined on a square lattice by the Hamiltonian:
\begin{eqnarray}
H=
-t \sum_{\langle i,j \rangle , \sigma} (c^{\dagger}_{i\sigma} c^{}_{j\sigma}
 +c^{\dagger}_{j\sigma} c^{}_{i\sigma}) - 
t_p \sum_{\langle\langle i,j \rangle\rangle , \sigma} (c^{\dagger}_{i\sigma} c^{}_{j\sigma}
 +c^{\dagger}_{j\sigma} c^{}_{i\sigma}) + 
U \sum_{i} n_{i \uparrow} n_{i \downarrow}
\label{glei1}
\end{eqnarray}
The number operator $n_{i\sigma}$ and the fermion creation operator $c^{\dagger}_{i\sigma}$
are defined as the ones with spin $\sigma$ at site $i$. 
$\sigma$ is the summation of spin up and spin down particles.
The summation with respect to
$\langle i,j \rangle$ is over the nearest neighbor  pairs, whereas the summation
with respect to $\langle\langle i,j \rangle\rangle$ is over the next nearest
neighboring pairs. Taking $t_p=0$ reproduces the usual single band Hubbard model \cite{Hub63}.
The hopping parameter $t=1$ is taken as the energy unit throughout this paper.

\section{Numerical Methods}

\subsection{Exact Diagonalization}

A basic technique to determine the ground state properties is the
exact diagonalization. 
Here one writes the Quantum problem as a eigenproblem
\begin{eqnarray}
H | \Psi \rangle = E | \Psi\rangle  
\end{eqnarray}
and searches for the lowest eigenvalue and eigenvector (\,eigenstate\,)
of the Hamilton-operator $H$.
We use for this paper the well known Lanczos-method \cite{CUL85}.

As basis states for the matrix-formalism of the Hamiltonian we use 
plane waves, i.\,e.\ we take 
basis-states $ |\Psi_i \rangle$ of the
form:
\begin{eqnarray}
|\Psi_i \rangle := c_{k_1 \up}^\dagger c_{k_2 \up}^\dagger \ldots c_{k_n
\up}^\dagger c_{k'_1 \down}^\dagger c_{k'_2 \down}^\dagger \ldots c_{k'_n
\down}^\dagger |0 \rangle,
\end{eqnarray}
where $|0\rangle $ is the vacuum state.
To reduce the amount of the basis states we apply the translation 
symmetry, and restrict the calculation to the subspace where the total momentum
 $K := \sum\limits_i (k_i + k'_i) = 0$.
As initial-vector for the Lanczos-procedure we use a random vector 
with the norm 1.

As Lanczos iteration we use \cite{GOL89}:
\begin{eqnarray}
H q_j = \beta_{j-1} q_{j-1} + \alpha_j  q_j + \beta_j q_{j+1} \qquad
\mbox{with} \qquad \beta_0 q_0 = 0
\end{eqnarray}

with $q_j$ being the Lanczos-vector in the $j$-th iteration.

The limitation of the exact diagonalization is the  
amount of  memory that is dramatically increasing with increasing 
system-size.
For a lattice with $N=n_x \cdot n_y$ points and  $n_e $ particles with 
spin up and $n_e$ particles with spin down the Hilbert-space has the dimension $D$:
\begin{eqnarray}
D= {N  \choose n_e}^2
\end{eqnarray}
This means, that a system with $N=4\cdot 4 = 16$ lattice-points and 
$n_e=5$ electrons per spin 
has the dimension $D={16 \choose 5}^2 = 4368^2 \approx 1.9 
\cdot 10^7$ and  can be 
solved with the Lanczos-method on state of the art computers. 
But a system with $N = 6\cdot 6 = 36$ and 
$n_e=13 $ electrons per spin direction has $D={36 \choose13}^2 \approx
(2.3108 \cdot 10^9)^2 \approx 5.3398 \cdot 10^{18}$ and can not be 
solved by exact diagonalization with
computer systems avaible today.

One method that circumvents some of these limitations is the stochastic
 diagonalization method.

\subsection{Stochastic Diagonalization}

For the stochastic diagonalization 
one also writes the 
physical quantum problem as an eigenproblem
to extract the ground state properties. But one does not use all 
basis-states\cite{RAE92}.
From the view of memory consumption
it would be possible to handle all system-sizes with this technique.
But due to a dramatic growth of the CPU-consumption with 
system-size it is so far
only possible to determine the Hubbard-model for systems up to $8\times8$ 
sites for small and intermediate interactions.

Sketching the basic ideas of the algorithm
we consider $n$ basis-states, which are orthonormal. The
matrix (\,Hamiltonian\,) for these $n$ basis-states is then denoted 
as $H^{(n)}$.
The problem is to
find the eigenvalues $E^{(n)}_i$. It is assumed, that they are ordered:
$E^{(n)}_1 \le E^{(n)}_2 \le \ldots \le E^{(n)}_n$. 
They can be obtained from $H^{(n)}$
with an unitary transformation $U^{(n)}$ 
\begin{eqnarray}
E^{(n)} = U^{(n)} H^{(n)} U^{{(n)}^T},
\end{eqnarray}
where $E^{(n)}$ is a diagonal matrix with the eigenvalues $E^{(n)}_1, 
E^{(n)}_2 , \ldots, E_n{(n)}$.
There are several standard algorithms for obtaining this unitary 
transformation.
In the classical Jacobi algorithm \cite{GOL89} for example the
transformation $U^{(n)}$ is a product of plane rotations $U^{(n,m)}(i_m,j_m)$.
\begin{eqnarray}
\label{EqUni}
U^{(n)} = \lim_{m \to \infty} \prod_{m} U^{(n,m)}(i_m,j_m).
\end{eqnarray}
Each plane rotation annihilates two off-diagonal elements
$H^{(n,m)}_{i_m,j_m}$ and $H^{(n,m)}_{j_m,i_m}$,
but also zeros in the off-diagonal elements can become nonzero after a
rotation. Only the sum
over all off-diagonal elements converges to zero. This means that,
 it is very important
for the convergence, in which order the rotations are performed.
Usually one always annihilates the largest off-diagonal element in the
following step. 
It is impossible to perform the infinite number of plane rotations in 
equation \ref{EqUni} with a computer, which is necessary for the exact 
diagonalization of $H^{(n)}$.
Therefore one only performs so many plane rotations, until
the absolute value of the largest
off-diagonal matrix elements is smaller than a threshold $T_R$.

In the modified Jacobi scheme \cite{RAE92} one only computes the 
smallest eigenvalue of $H^{(n)}$
i.\,e.\ one keeps 
$i_m $ to be $1$. This has the advantage, that it is not necessary 
to store the whole matrix and to make less rotations. 
One can prove \cite{RAE92} that the 
diagonal-element $H^{(n)}_{1,1}$ is the smallest eigenvalue of the matrix $H^{(n)}$, 
if the rotations are 
performed as described in \cite{RAE92}.

A short description of the algorithm is given by:
\begin{enumerate}
\item choose one basis state as an initialization.
\item generate a set of new trial-states.
\item search the best trial-state for expanding the 
($n\times n$)-matrix to an  $((n+1)\times (n+1))$-matrix.
The best trial-state, is the state, for which the energy reduction 
$\Delta_m$ with the unitary transformation $U^{(n+1,1)}(1,n+1)$ 
is the largest.
\item if the reduction $\Delta_m$ of the best state is bigger than a
threshold $T_A$, than this best state is added to the matrix and we apply
plane rotations $U^{(n+1,m)}$ (\,$m \ge 1$\,) to the
matrix $H^{(n+1)}$ until the size of all off-diagonal elements
$H^{(n+1,m)}_{1,j}$ with $j=2, \ldots, (n+1)$ is smaller than a 
given threshold $T_R$.
\item if the reduction $\Delta_m$ of the best trial-state is smaller
than the threshold $T_A$ reduce the threshold $T_A$.
\item if the convergence criteria are satisfied stop the iterations.
\item goto step 2.
\end{enumerate}

One problem of the algorithm is: what are good convergence criteria?
One is the convergence of the ground state energy $E^{(n)}_1$ 
(\,the smallest eigenvalue 
of the matrices $H^{(n)}$\,) with increasing number of states $n$.
Others are the convergence of the physical properties one is 
interested in, like correlation-functions.
An additional good criterion for the convergence of the algorithm is the norm
$N^{(n,m)}_1$ of the first column of the matrix $H^{(n,m)}$:
\begin{eqnarray}
\label{Eqnorm}
N^{(n,m)}_1 := \sqrt{\sum_{i=1}^n {H^{(n,m)}_{i,1}}^2}
\end{eqnarray}

\subsection{The Projector Quantum Monte Carlo Method}

The QMC-algorithm used in this paper to investigate the ground state
properties of the Hubbard model is the Projector-Quantum-Monte-Carlo (PQMC)
algorithm. The basic idea of this algorithm is to use a projection operator
$e^{-\theta H}$ to extract the ground state $\vert \Psi_0 \rangle$
of the Hamiltonian $H$ from the trial wave function $\vert \Psi_T \rangle$ \cite{KOO82}:
\begin{eqnarray}
\vert \Psi_0 \rangle = \lim_{\theta \to \infty} e^{-\theta H} \vert \Psi_T \rangle
\end{eqnarray}
The trial wave function $\vert \Psi_T \rangle$ is chosen to be a Slater determinant.

Sorella et al.\ \cite{Sor89} have developed a stable 
algorithm for this purpose.
We give a short description to introduce the notation.
The Hubbard Hamiltonian (\ref{glei1}) then reads  as :
\begin{eqnarray}
H&=&H_{kin}+H_{hub} \\\nonumber
H_{kin}&=&
-t \sum_{\langle i,j \rangle , \sigma} (c^{\dagger}_{i\sigma} c^{}_{j\sigma}
 +c^{\dagger}_{j\sigma} c^{}_{i\sigma}) - 
t_p \sum_{\langle\langle i,j \rangle\rangle , \sigma} (c^{\dagger}_{i\sigma} c^{}_{j\sigma}
 +c^{\dagger}_{j\sigma} c^{}_{i\sigma}) \\\nonumber
H_{hub}&=&
U \sum_{i} n_{i \uparrow} n_{i \downarrow}
\end{eqnarray}

Using the Trotter-Suzuki formula \cite{Suz76} the exponential of the
Hubbard Hamiltonian is rewritten as:
\begin{eqnarray}
e^{-\theta (H_{kin}+H_{hub})} = \lim_{m \to \infty} \left( e^{-\frac{\theta}{m}H_{kin}}
e^{-\frac{\theta}{m}H_{hub}}\right)^m
\end{eqnarray}

For the interaction term $H_{hub}$ we introduce the discrete Hubbard Stratonovich transformation
 \cite{Hir83} to eliminate the quartic term as:

\renewcommand{\arraystretch}{1.5}
\begin{eqnarray}
e^{-\tau H_{hub}} &=& \prod_{i} e^{-\tau Un_{i \uparrow} n_{i\downarrow}} \\
                &=&
\sum_{\sigma =\pm 1} e^{-\frac{\tau}{2} \sum_i \hat{n}_i U} e^{\lambda \sum_{i} \sigma_{i}
\hat{m}_i} \\
&\equiv& \sum_{\sigma=\pm 1} e^{h_{hub}(\sigma)} \quad .
\end{eqnarray}

where the following definitions have been used:
\begin{eqnarray}
\hat{n}_i\equiv
 \left\{
        \begin{array}{lll}
         n_{i\uparrow}+n_{i\downarrow}-1 & \quad \mbox{if} & U \le 0 \\
          n_{i\uparrow}+n_{i\downarrow} & \quad \mbox{if} & U > 0
         \end{array}
         \right. \quad \mbox{and} \quad \quad
\hat{m}_i\equiv
         \left\{
         \begin{array}{lll}
         n_{i\uparrow}+n_{i\downarrow}-1 & \quad \mbox{if} &  U \le 0 \\
          n_{i\uparrow}-n_{i\downarrow} &  \quad \mbox{if} &  U > 0
          \end{array}
          \right. 
\end{eqnarray}
and
\begin{eqnarray}
  \cosh(\lambda) & \equiv & e^{\frac{\tau}{2}\vert U \vert } \\
             \tau & = &   \frac{\theta}{m}\quad .
\end{eqnarray}
\renewcommand{\arraystretch}{1.0}
The elements of the discrete Hubbard Stratonovich field $\sigma$ 
can only take the values $\sigma_i = \pm 1$.
The Hubbard Stratonovich field is often referred to as 
Ising-spins or $\sigma$-spins.

The projection of the ground state then reads as:

\begin{eqnarray}
e^{-\theta H} \vert \Psi_T\rangle &=& \sum_{\sigma} \prod_{l=1,m} e^{-\frac{\theta}{m}H_{kin}} 
e^{h_{hub}(\sigma(l))} \vert \Psi_T\rangle \\\nonumber
             &\equiv&  \sum_{\sigma} \vert \sigma \rangle 
\end{eqnarray}

The expectation value of an observable $A$ in the ground state is now
given by:

\begin{eqnarray}
\langle\langle A \rangle\rangle  =
\frac{\sum\limits_{\sigma \sigma'}\langle \sigma \vert  A \vert \sigma' \rangle}
{\sum\limits_{\sigma \sigma'}\langle \sigma \vert \sigma' \rangle} \quad .
\end{eqnarray}
Normalizing the expectation value $\langle \sigma \vert  A \vert\sigma' \rangle$
with $\langle \sigma \vert \sigma' \rangle$ and using the following definitions
 \setlength{\jot}{7.0pt}
\begin{eqnarray}
\langle A \rangle &=& \frac{\langle \sigma \vert A \vert \sigma'\rangle}
{\langle \sigma \vert \sigma'\rangle} \label{gla15}\\
w(\sigma,\sigma')&=&\frac{\vert \langle \sigma \vert \sigma' \rangle \vert}
{\sum\limits_{\sigma\sigma'} \vert\langle \sigma \vert \sigma' \rangle\vert} \quad .
\end{eqnarray}
the expectation value reads as:
\begin{eqnarray}
\langle\langle A \rangle\rangle =
\frac{\sum\limits_{\sigma \sigma'} w(\sigma,\sigma') {\rm sign}\Big(\langle\sigma\vert\sigma'\rangle\Big)
\langle  A \rangle}
{\sum\limits_{\sigma \sigma'} w(\sigma,\sigma') {\rm sign}\Big(\langle\sigma\vert\sigma'\rangle\Big)} \quad .
\end{eqnarray}

The sum over $\sigma,\sigma'$ is now calculated by an importance sampling 
Monte-Carlo procedure. The algorithm can be sketched in the following way:
\begin{enumerate}
\item create a new configuration $(\sigma,\sigma')_{n+1}$ from
the old configuration $(\sigma,\sigma')_{n}$ by an elementary move.
\item calculate the transition probability 
\begin{eqnarray}
P\Big((\sigma,\sigma')_{n} \to (\sigma,\sigma')_{n+1}\Big)=\frac{w(\sigma,\sigma')_{n+1}}{
w(\sigma,\sigma')_{n}}
\end{eqnarray}
\item calculate a random number $z$ between 0 and 1
\item if $z<P\Big((\sigma,\sigma')_{n} \to (\sigma,\sigma')_{n+1}\Big)$ accept the
new configuration $(\sigma,\sigma')_{n+1}$ else keep the old 
configuration $(\sigma,\sigma')_n$
\item with this procedure one generates the Monte Carlo configurations 
with the 
probability $w(\sigma,\sigma')$. The expectation value of an observable $A$ is 
then given by:
\begin{eqnarray}
\langle\langle A \rangle\rangle=\frac{\sum\limits_{\sigma \sigma'} 
{\rm sign}\Big(w(\sigma,\sigma')\Big)\langle  A \rangle }
{ \langle {\rm sign} \rangle } 
\end{eqnarray}
with the average sign $\langle {\rm sign} \rangle$:
\begin{eqnarray}
\label{Eqsign}
\langle {\rm sign} \rangle = 
\sum\limits_{\sigma \sigma'}{\rm sign}\Big(w(\sigma,\sigma')\Big)
\end{eqnarray}
\end{enumerate}

In order to use the determinant $\langle \sigma \vert \sigma' \rangle$ 
as a probability one has to take the absolute value. The sign of this
determinant then becomes part of the expectation value which 
causes the so called Minus-Sign problem. 

We implemented three different kinds of elementary moves, acting on the
$N\times m$ array of the $\sigma$-configurations ($\sigma$-spins), where
$N$ is the system-size and $m$ the number of Trotter slices. 
We are motivated by results obtained for QMC-Simulations
of the Spin $\frac{1}{2}$ Heisenberg model where evidence 
was presented that the elementary move
influences the convergence of the algorithms and can even lead to 
wrong results \cite{Rae84}.
\begin{enumerate}
\item The Cluster move flips a certain number $S$ of randomly chosen $\sigma$-spins.
The criteria for the choice of the number $S$ was, that the acceptance rate 
of these moves was at least 40\%. 
\item The Slice move flips a certain number $S$ of randomly chosen $\sigma$-spins
only on a single Trotter slice. By visiting 
the slices in consecutive order and storing intermediate results of the calculation the
amount of computations can be reduced significantly. Details of the
implementation will be given elsewhere~\cite{tobepublished}.
\item The Single Spin elementary move only flips one randomly chosen single $\sigma$-spin. This
allows a significant simplification of the algorithm \cite{Ima89}.
\end{enumerate}

For the efficient implementation of the algorithm the elementary move is more than 
a minor detail and the execution time depends significantly on  
the choice of the elementary move~\cite{tobepublished}.

QMC simulations of the Hubbard model are in general very time consuming.
The calculations were carried out on a parallel computer. Details of
the parallelized algorithm have already been published
 elsewhere \cite{parallel}.

Finally we define two important quantities, the Monte-Carlo steps and the
error bars in the QMC algorithm. One Monte Carlo step (MCS) is completed when 
the program has tried to flip all $N \cdot m $ $\sigma$-spins (\,$N$ is the 
size of the lattice and $m$ is the number of Trotterslices\,). 

If after one run through the lattice the acceptance rate is too
low, additional runs are performed until the acceptance rate
is reached.
Second the error bars are 
taken over bins, i.e. typically 50 measurements are sampled in one 
average $\bar{A}$. The error bar is now calculated as the fluctuations of 
these averages $\bar{A}$.

\section{Observables}

As a check whether the different programs are correct we compare different
observables. The definitions of these observables will be presented in this
section. 

The observable  commonly used for the check of a correct ground state
determined with a
method is the ground state energy per lattice site $E_0/N$.

A common way to study superconductivity in the Hubbard model is to
examine the two particle correlation functions for the  occurrence of Off Diagonal 
Long Range Order (\,ODLRO\,) in the model \cite{YAN62}.
In order to do this we calculate the full correlation function
with d$_{x^2-y^2}$ symmetry \cite{hus94}
\begin{eqnarray}
\label{Eqfull}
C^{Full}_{d_{x^2-y^2}}(r)= \frac{1}{N}
\sum_{i,\delta,\delta'}
g^{}_{\delta}g^{}_{\delta'}\langle c^{\dagger}_{i\uparrow}
c^{\dagger}_{i+\delta\downarrow} c^{}_{i+\delta'+r \downarrow}
c^{}_{i+r \uparrow} \rangle
\end{eqnarray}
that measures superconducting correlations as a function of 
lattice vector $r$.
The phase factors $g^{}_{\delta},g^{}_{\delta'}$ are $1$ in 
$x$-direction and $-1$ in $y$-direction. The sum over $i$ goes over the whole
lattice. The sums with respect to $\delta$ and $\delta'$ are the independent 
sums over the nearest neighbors of $i$.

When studying the superconducting properties of small Hubbard-Clusters the full
correlation function is not a very appropriate measure as this expectation value
also contains contributions from the one-particle Greens functions 
$C_\sigma^{single} (r) = \frac{1}{N} \sum_i 
\langle c^{\dagger}_{i\sigma} 
c^{}_{i+r\sigma} \rangle$. In principle these
one particle contributions can be neglected as they decrease to zero
with increasing distance $|r|$. 
But when studying small clusters their influence on the results has to be taken into account~\cite{Whi89}.
Therefore we study the vertex correlation function
\begin{eqnarray}
\label{Eqvertex}
C^{Vertex}_{d_{x^2-y^2}}(r)& = & C^{Full}_{d_{x^2-y^2}}(r)- 
\sum_{\delta, \delta'} \left( g_\delta g_\delta' C_\uparrow^{single} (r)
C_\downarrow^{single} (r+\delta'-\delta) \right) \\\nonumber
&=& \frac{1}{N}
\sum_{i,\delta,\delta'} \left(
g^{}_{\delta}g^{}_{\delta'}\langle c^{\dagger}_{i\uparrow}
c^{\dagger}_{i+\delta\downarrow} c^{}_{i+\delta'+r \downarrow}
c^{}_{i+r \uparrow} \rangle
- g^{}_{\delta}g^{}_{\delta'}
\langle c^{\dagger}_{i\uparrow}
c^{}_{i+r \uparrow} \rangle \langle c^{\dagger}_{i+\delta\downarrow}
c^{}_{i+\delta'+r \downarrow} \rangle \right) \quad .
\end{eqnarray}

\section{Comparison}

In this section we compare the results obtained with 
ED, SD and PQMC. 
We restrict ourselves to the case of the $4\times 4$ system. This is 
the largest square
lattice that can be solved with the exact diagonalization. 
The goal is to check the correct convergence of the QMC-algorithms.
By this we mean for the PQMC, that we check, whether
the Monte Carlo procedure converges correctly and whether
the projection parameter $\Theta$ and the
Trotter decomposition $m$ are chosen sufficiently large to obtain the 
true ground state.
Special emphasis lies on various technical problems that come with the 
application of the QMC method as there are 
the minus sign problem, numerical instabilities and  
statistical fluctuations. 
In the case of the SD we investigate whether the number of basis states
accumulated is sufficient to be a good approximation of the "true"
ground state.

Inspired by the results for the RHM \cite{hus94} for larger lattice
sizes we first turn our attention to the RHM.
The most common indicator for reaching the ground state is the 
ground state energy per lattice site $E_0/N$.
In table~\ref{tab1} and~\ref{tab2} we present a comparison of 
the ground state energies of the ED, SD and the three different 
implementations of the PQMC-algorithm (\,section II.C\,). 
The error bars on the QMC results can be estimated to be $0.25$\%. The result of ED, SD
and  QMC are the same within these error bounds.
We find excellent agreement between these different methods for the 
ground state energy
even when a minus sign problem is present in the calculations. 
This can be seen from
table~\ref{tab3} where we list the average sign 
(\,$\langle {\rm sign} \rangle$, eq.\ \ref{Eqsign}\,) for the
simulations presented in table~\ref{tab1} and~\ref{tab2}.
The $\langle {\rm sign} \rangle$ has the same value in all
three PQMC-algorithms we used.
The SD method gives in contrast to the PQMC algorithm an upper 
bound for the exact ground state energy $E_0$. Therefore comparing two
energies obtained with SD it is clear, that the lower
energy is closer to the exact energy. 
For the PQMC the error bar gives a bound for the ground state energy, but
it is not possible to deduce from the PQMC value whether the energy is
above or below the exact energy.

It has turned out that comparing ground state energies is not a 
very sensitive indicator whether the
PQMC has reached the true ground state~\cite{MIC96}, ~\cite{FET96}.

Our interest lies on the expectation values of the superconducting correlation
function for the RHM in the $d_{x^2-y^2}$-channel.

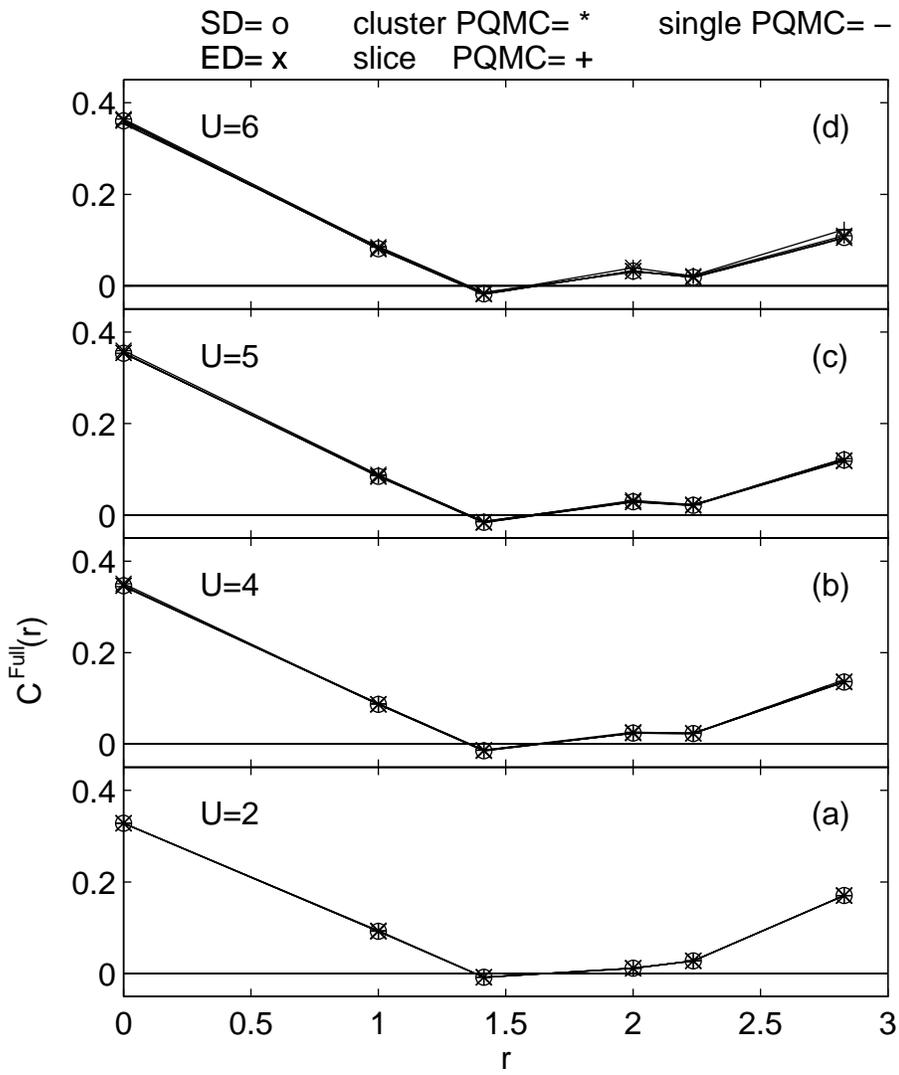
\begin{figure}
\caption{Comparison of the full correlation function with $d_{x^2-y^2}$ symmetry
$C^{Full}_{d_{x^2-y^2}}(r)$ for a $4\times 4$ Hubbard cluster with $N_e=5+5$,
$t_p=-0.22$ and $U=2$, $4$, $5$ and $6$. (\,$\Theta =8$, $\tau = 0.125$\,)}
\label{fig1}
\end{figure}

In figure~\ref{fig1} we present the comparison of $C^{Full}_{d_{x^2-y^2}}(r)$ 
(eq.~\ref{Eqfull}) for 
five different methods namely the ED, SD, single-PQMC, slice-PQMC and cluster-PQMC.
For a $4\times 4$ Hubbard system with $t_p=-0.22$, $N_e=5+5$ and $U=2$,
$4$, $5$ and $6$ respectively
we observe perfect agreement of all five programs. 
The points for all five methods fall on identical positions and cannot be 
distinguished.

As we outlined in the previous section the more appropriate indicator for the
existence of ODLRO in small clusters is the vertex correlation function 
$C^{Vertex}_{d_{x^2-y^2}}(r)$ 
(eq.~\ref{Eqvertex}).
The absolute values of this vertex correlation function compared to the
full correlation function  turned out to be
very small for the RHM~\cite{hus94} giving rise to the question whether these small results
might be artifacts of the various problems that come with the application
of the QMC-method.

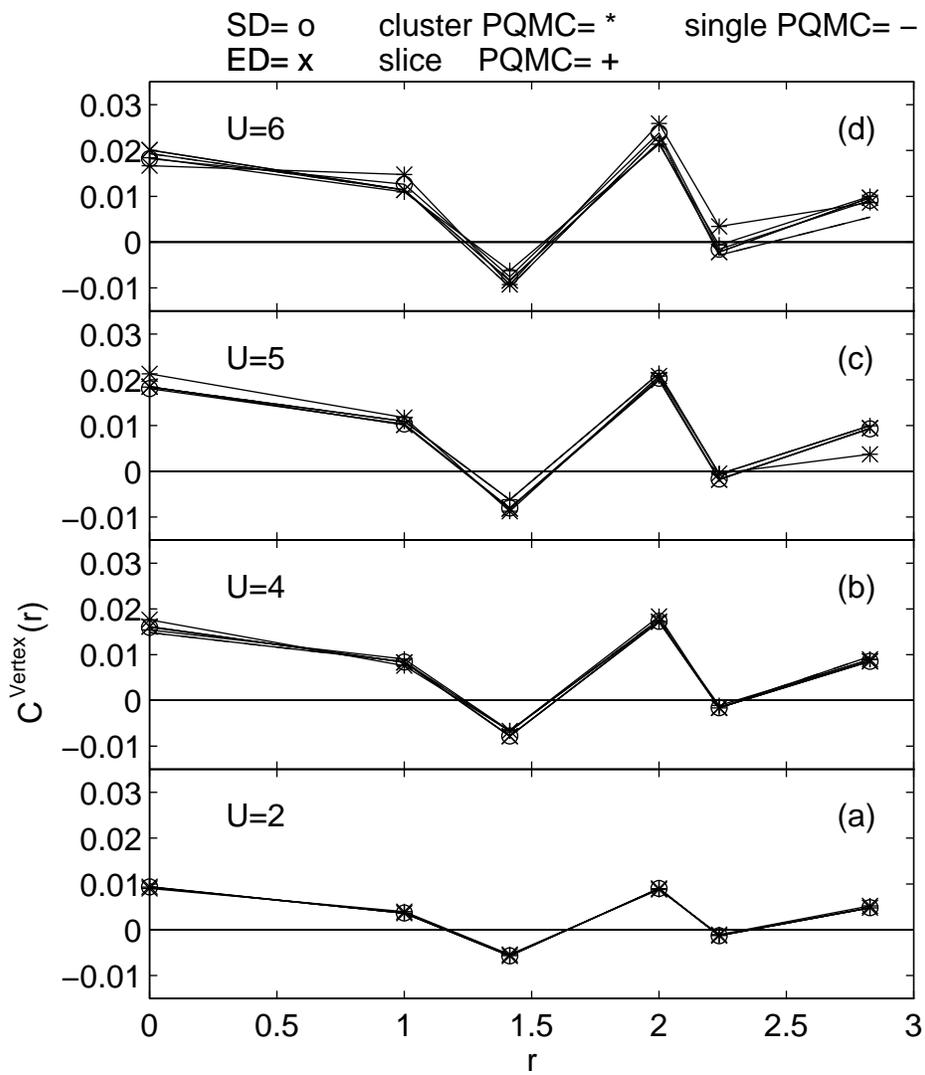
\begin{figure}
\caption{Comparison of the vertex correlation function with $d_{x^2-y^2}$ symmetry
$C^{Vertex}_{d_{x^2-y^2}}(r)$ for a $4\times 4$ Hubbard cluster with $N_e=5+5$,
$t_p=-0.22$ and $U=2$, $4$, $5$ and $6$. (\,$\Theta =8$, $\tau = 0.125$\,)}
\label{fig2}
\end{figure}

In figure~\ref{fig2} a-d we present a systematic investigation of 
the $4\times 4$-system with $5+5$ electrons for various values of $U$. 
In contrast to the case of the full correlation function $C^{Full}_{d_{x^2-y^2}}(r)$
we observe that in the case of the vertex correlation function 
$C_{d_{x^2-y^2}}^{Vertex}$ (eq.~\ref{Eqvertex}) only simulations with
$U=2$ and $U=4$
show again a perfect agreement. For larger interactions $U=5$ and $U=6$ we find
strong fluctuations of the different PQMC methods coinciding with the occurrence of a minus
sign problem in the simulations (table~\ref{tab3}).
Thus we have shown that whereas the ground state energy and the full correlation function
seem perfectly converged this is no proof that the vertex correlation function
is converged in these runs.
The fluctuations in the correlation functions have increased dramatically
with interaction $U$ despite the significant higher number of MCS,
as shown in table \ref{tabMCSpos}.

We now turn our attention to the AHM. Here we consider both the cluster
and the single spin dynamics in the PQMC algorithm. The AHM shows 
superconductivity in the on-site s-wave channel. In this case there
is no summation over the nearest neighbors, i.e.\ no sum with respect to 
$\delta$ and $\delta'$ in the eq.\ \ref{Eqfull} and \ref{Eqvertex}.
Of course the phasefactors $g_\delta = g_{\delta'} $ are equal 1.

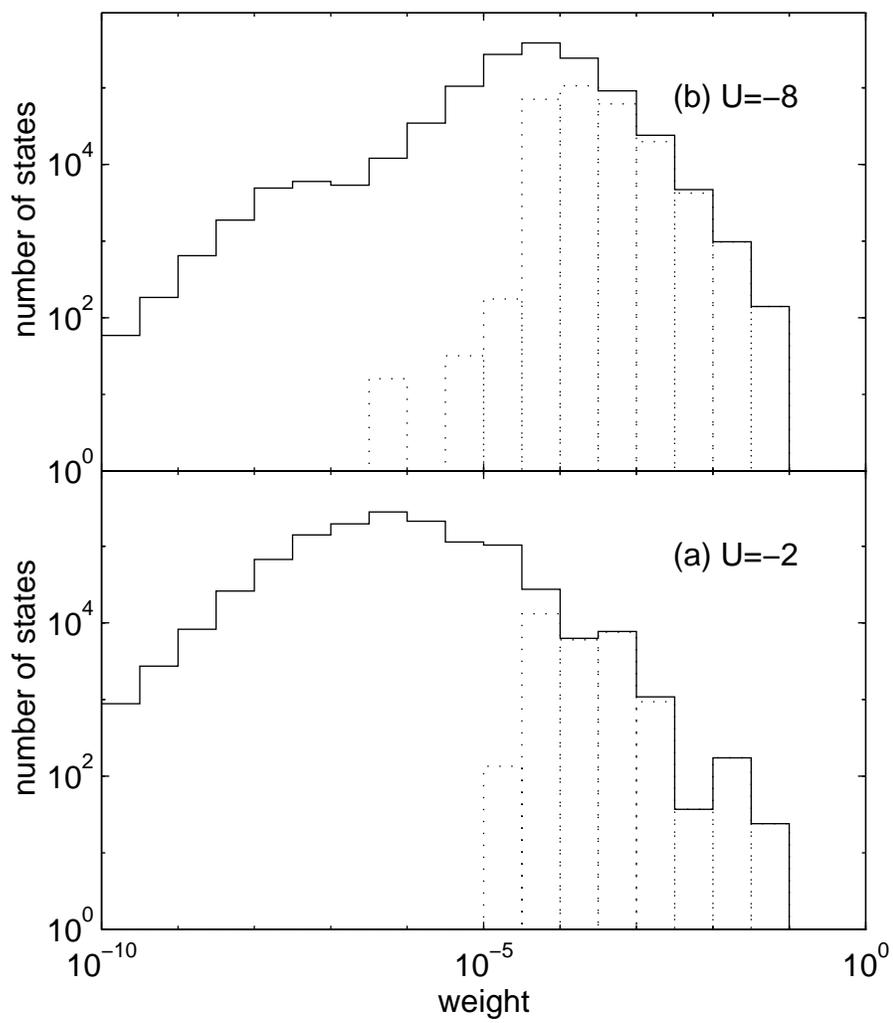
\begin{figure}
\caption{Comparison of the weights of the basis states in the exact 
diagonalization (ED) and the stochastic diagonalization (SD) in a 
$4\times 4$ system with $N_e = 5+5$ electrons for the interaction
$U=-2$ and $U=-8$ and $t_p= 0.0$.}
\label{figweights}
\end{figure}

First we compare the ground state energy for PQMC  with ED and SD (\,table
\ref{tab5}\,). 
Note that for the SD deviations from 
the ED values develop with increasing interaction strength $|U|$. 
For interactions $U=-6$ and $U=-8$ the 
SD algorithm has convergence problems. For these high values of $|U|$
the k-space representation  of the basis states is no longer appropriate. 
This can be seen from figure \ref{figweights}.
The SD algorithm is able to find 
almost every important state (i.e.\ all states with a large weight)
for the interaction $U=-2$. In the interval $0.0001$ to $0.00032$
there are $6276$ states in the ground state of the exact diagonalization. 
$6012$ states are in the same interval in the basis of important 
states of the SD. For $U=-2$ the SD has collected almost every state 
with a weight larger than $10^{-4}$. 

The number of states with a relatively large weight increases
for  stronger interactions $|U|$. In the interval $0.001$ to $0.0032$ 
the number of states in the ground state of the exact diagonalization 
goes from $1\,082$ ($U=-2$) to $24\,003$ ($U=-8$)
(fig. \ref{figweights}).
Therefore more states are necessary as a good approximation of
the ground state in the SD.
These convergence problems can be demonstrated even clearer in 
table \ref{tabstates}.
This table shows additionally to the number of states $N_{SD}$ in the 
basis of the SD (\,plotted in fig.\ \ref{figweights}\,)
the norm $N_1^{(n,m)}$ of the first column 
(\,eq. \ref{Eqnorm}\,) and the threshold $T_A$ for the acceptance of new
states.  
Because the CPU-time is limited  \cite{FET96}, \cite{MIC96} 
in practice the number of states $N_{SD}$
can not exceed to some hundred thousand. 
From table \ref{tabstates} one can see, that  $N_{SD}$ and
$N_1^{(n,m)}$ as well as $T_A$ are increasing with $|U|$. 
For practical purposes the SD is not able to find a good approximation
of the exact ground state beyond a certain interaction.
We were not able to resolve these problems by the use of a real space 
representation of the basis states \cite{unpublished2}.

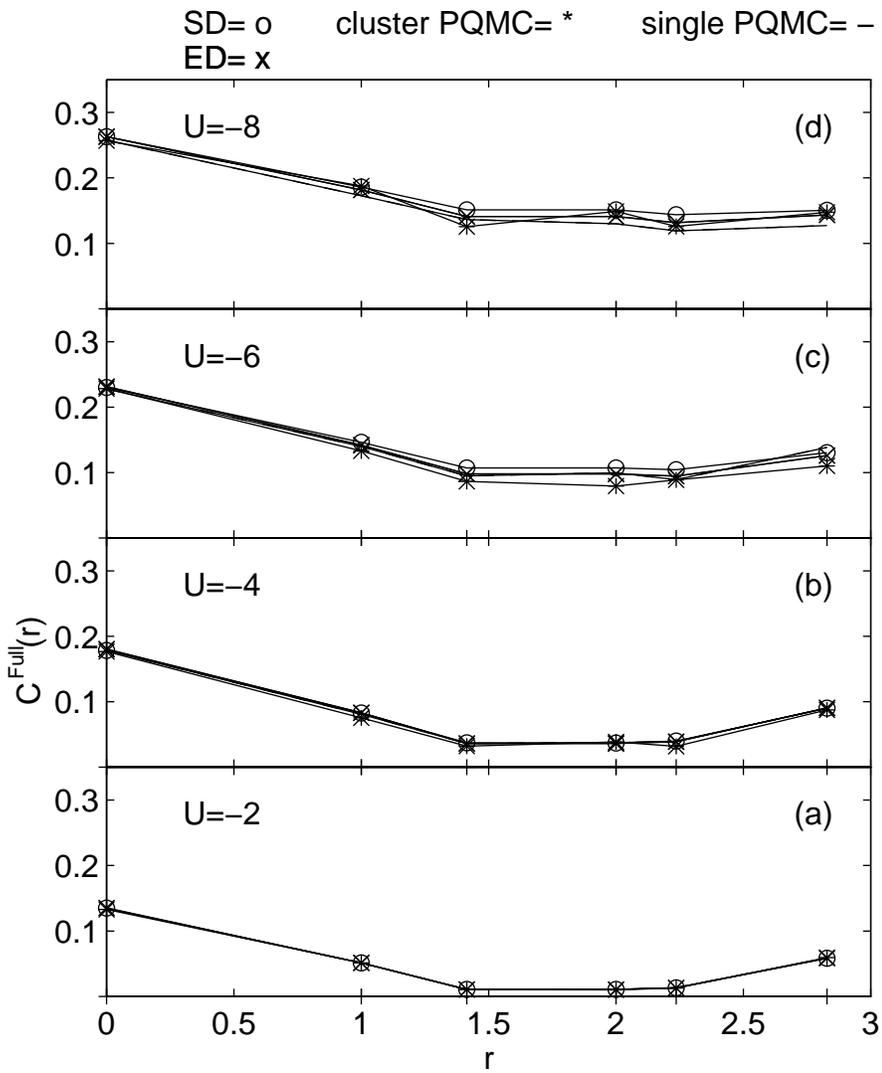
\begin{figure}
\caption{
\label{fig3}
Comparison of the full correlation function with on-site s-wave 
 symmetry
$C^{Full}_{os}(r)$ for a $4\times 4$ Hubbard cluster with $N_e=5+5$,
$t_p=0$ and $U=-2$,$-4$, $-6$ and $-8$. (\,$\Theta =8$, $\tau = 0.125$\,)}
\end{figure}

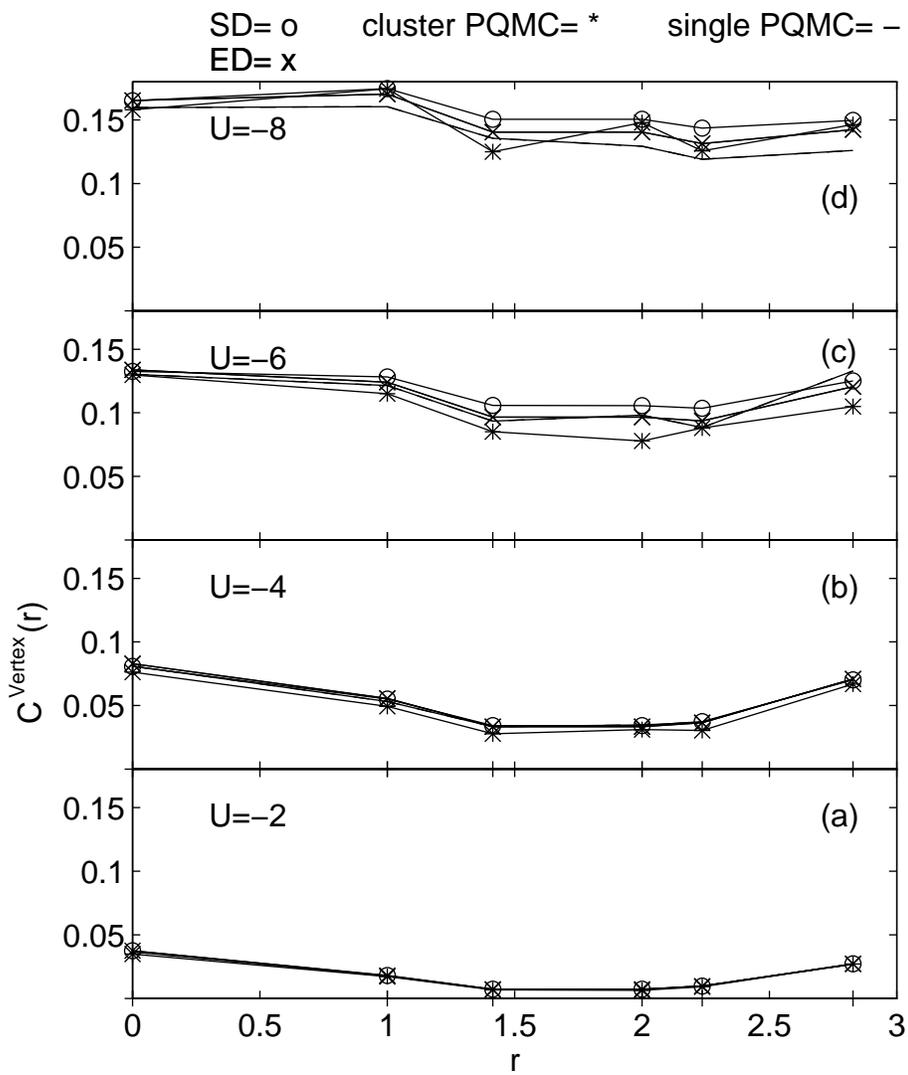
\begin{figure}
\caption{
\label{fig4}
Comparison of the vertex correlation function with on-site s-wave 
symmetry
$C^{Vertex}_{os}(r)$ for a $4\times 4$ Hubbard cluster with $N_e=5+5$,
$t_p=0$ and $U=-2$, $-4$, $-6$ and  $-8$. (\,$\Theta =8$, $\tau = 0.125$\,)}
\end{figure}

As a second step we compare the superconducting correlation function 
with on-site s-wave symmetry for the AHM.
Figure \ref{fig3} shows $C_{os}^{Full}$ and \ref{fig4} shows 
$C_{os}^{Vertex}(r)$
for $U=-2$, $-4$, $-6$ and $-8$ for a $4\times4$ system with $t_p =0$. 
The comparison of the vertex correlation function with the Lanczos 
method clearly indicates that for $U=-6$ and especially for $U=-8$ 
the cluster program shows deviations from the ED values. 
These deviations are comparable to the case of the RHM, where we 
also observe convergence problems for increasing interaction $U$.
We tried to find out where these convergence problems might come from.
On the one hand we investigated the dependence on the projection
paramter $\Theta$. In figure \ref{fig8} and \ref{fig9} we show
the dependence of the cumulated plateau of the vertex correlation
function $C_{os}^{Vertex}(r)$ and of the ground state energy per site 
$E_0/N$ of the projection parameter $\Theta$.
It clearly indicates that a projection parameter of $\Theta =8 $ as
it was used throughout the rest of the paper is sufficient to
reach the ground state. Note that especially the ground state energy does
not depend very strongly on the projection parameter.

Furthermore we investigated whether numerical instabilities might cause
the convergence problems. In table \ref{tab10} we compare the
ground state energy for different numbers of stabilizations $nstab$.
To reduce numerical instabilities the product of $m$ imaginary time slices
is being stabilized by a modified Gram-Schmidt orthogonalization 
\cite{Sor89}.

A stabilization is performed every $nstab$ slices of the
Trotter decomposition. The results agree perfectly so we
conclude that the simulations were numerically stable.

Finally we checked whether the MC process depends on the starting seed
of the random number generator. Figure \ref{fig10} shows
the ground state energy
for different seeds. Here we also cannot discover any unusual behaviour.

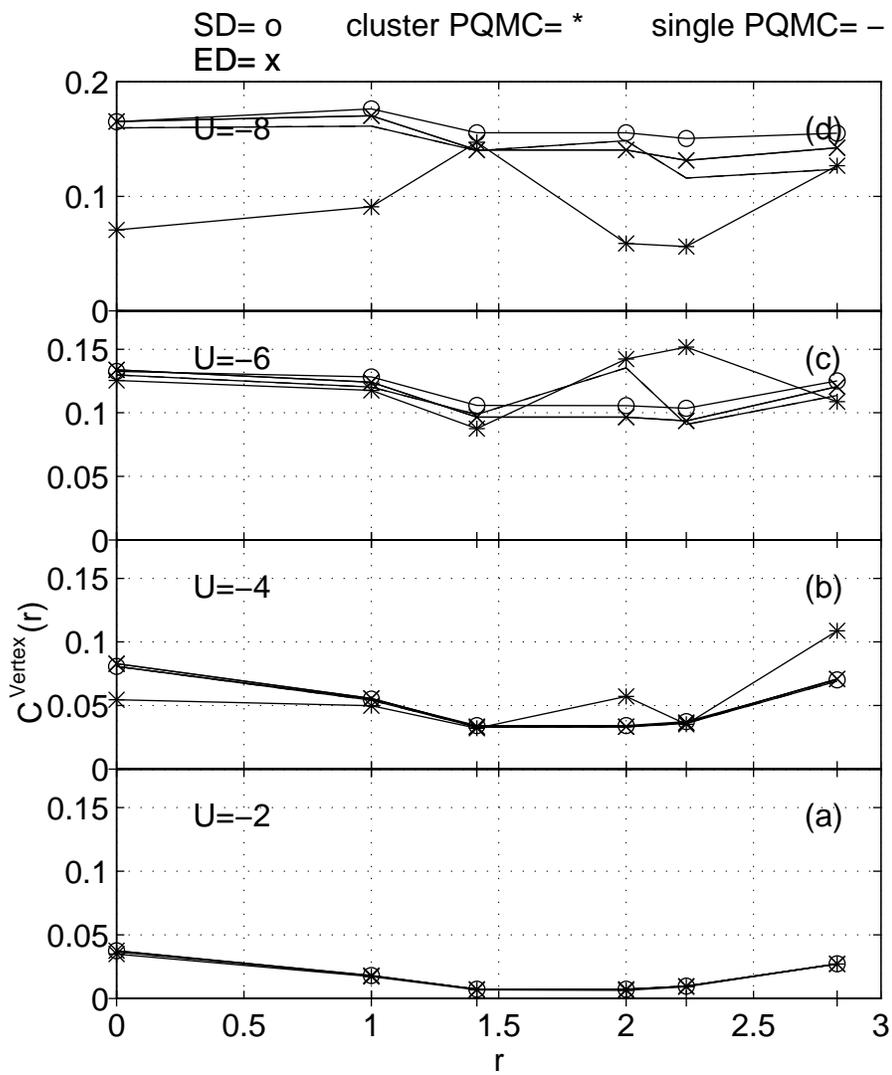
\begin{figure}
\caption{
\label{fig7}
Comparison of the vertex correlation function with on-site s-wave  
symmetry
$C^{Vertex}_{os}(r)$ for a $4\times 4$ Hubbard cluster with $N_e=5+5$,
$t_p=0$ and $U=-2$, $-4$, $-6$ and $-8$. (\,$\Theta =8$, $\tau = 0.125$\,),
The runs are unconverged meaning that for $|U|>-2$ the number of Monte Carlo 
Steps was chosen significantly lower than in figure 5. Details are given in
table X}
\end{figure}

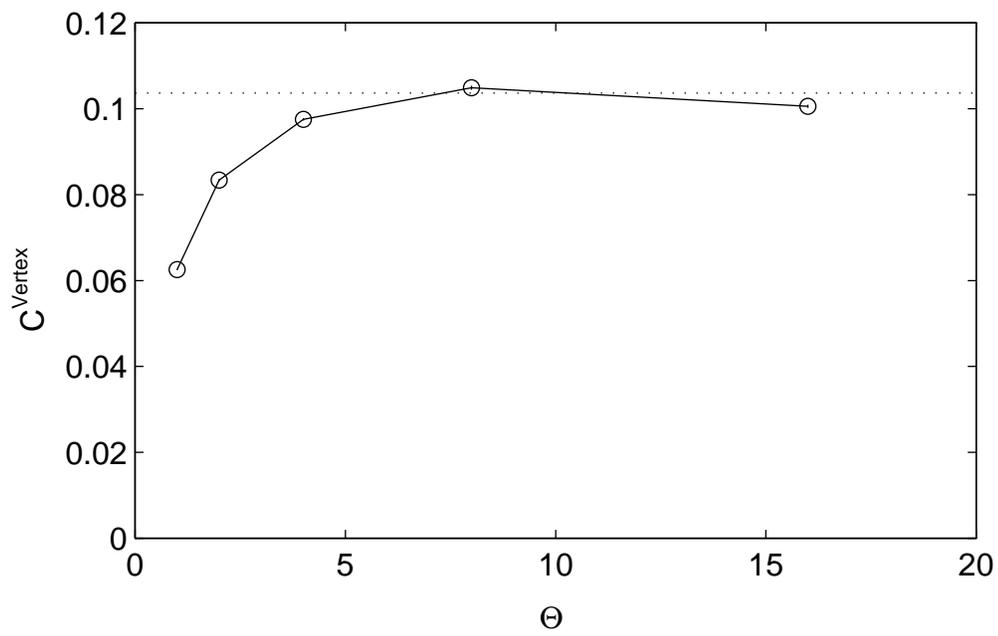
\begin{figure}
\caption{
\label{fig8}
Investigation of the behaviour of the vertex correlation function with on-site s-wave  
symmetry
$C^{Vertex}_{s_{os}}(r)$ for a $4\times 4$ Hubbard cluster with $N_e=5+5$,
$t_p=0$ and $-6$ for $\Theta=$ 1,2,4,8 and 16. ($\tau=0.125$)
Dashed line is the value of the exact diagonalization.}
\end{figure}

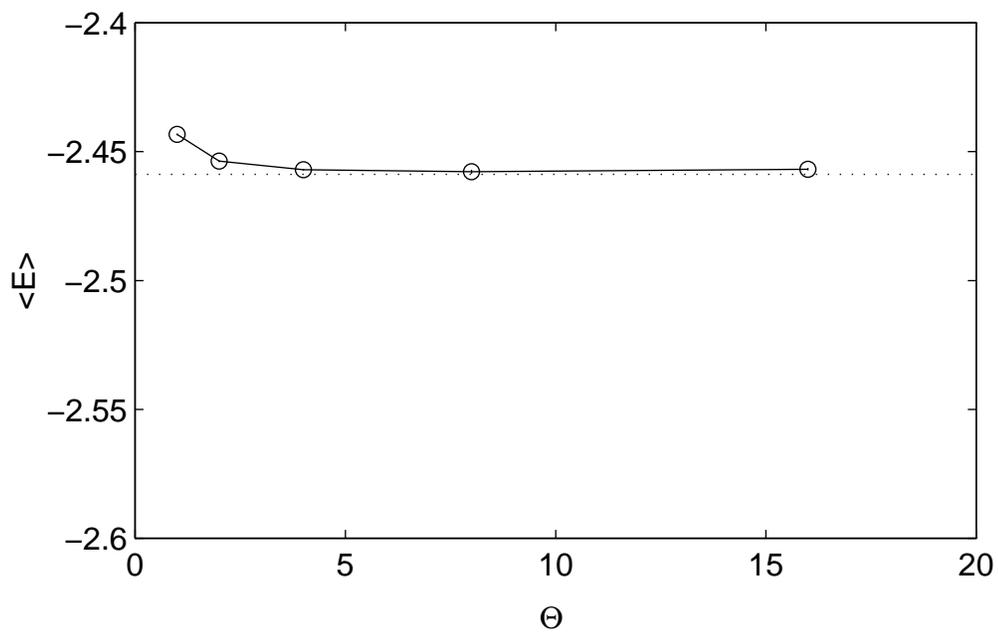
\begin{figure}
\caption{
\label{fig9}
Investigation of the behaviour of the ground state energy 
for a $4\times 4$ Hubbard cluster with $N_e=5+5$,
$t_p=0$ and $-6$ for $\Theta=$ 1,2,4,8 and 16. ($\tau=0.125$)
Dashed line is the value of the exact diagonalization.}
\end{figure}

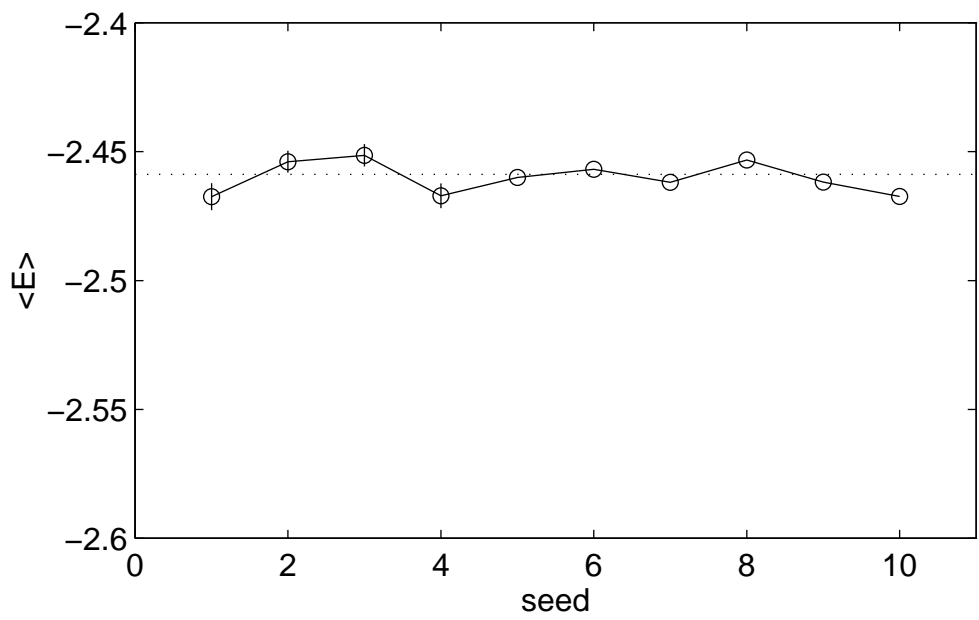
\begin{figure}
\caption{
\label{fig10}
Investigation of the behaviour of the ground state energy 
for a $4\times 4$ Hubbard cluster with $N_e=5+5$,
$t_p=0$ and $U=-6$ for different seeds of the random number generator.
($\Theta=8$, $\tau=0.125$) Dashed line is the value of the exact 
diagonalization and the full line is to guide the eyes.}
\end{figure}

Thus even in the absence of a minus sign problem for the AHM the 
PQMC algorithms have convergence problems. An even more striking result 
is seen comparing table \ref{tabMCSpos} and \ref{tabMCS}. The number 
of MCS needed for convergence
both in the RHM as well as in the AHM increase with interaction strength
$|U|$ for both the single and the cluster algorithm in the same 
manner. As a comparison we show in fig.~\ref{fig7} the results for 
unconverged runs.
The corresponding number of MCS (PQMC) and number of basis state (SD) 
are listed in table \ref{tabMCSunconverg}.

Given only limited CPU power, we come to the result, that even when 
there is no minus sign problem as for the example in the case of 
the PQMC algorithm for the attractive Hubbard model or of
the SD the presented numerical algorithms cannot handle  
values of $U$ beyond  a critical interaction, 
because of convergence problems.

\section{Conclusion}

We presented a detailed comparison of the results of exact 
diagonalization, stochastic diagonalization and  the PQMC. 
For the PQMC we implemented three different elementary moves, namely
the single spin flip move, the slice move and the cluster move.
We showed that there is perfect agreement
between all five methods for the ground state 
energies and the full correlation function for the RHM and the AHM.  
Also for  the  vertex correlation functions, that are
very small in magnitude in the repulsive case,  the exact 
diagonalization results can be  confirmed by PQMC 
and stochastic diagonalization. But for large interactions it is 
necessary to perform dramatically more MCS in the PQMC respectively 
to collect much more states in the SD.
The observation that with an increase of $|U|$ 
the amount of CPU-time is enhanced dramatically leads
to the problem that these calculations cannot be performed
with state of the art computers for large interaction strengths $|U|$.
This increase in CPU-time was found both, in the RHM where it was
expected due to the occurrence of the well-known minus sign problem
and in the AHM that is minus sign free.

\section{acknowledgment}

We thank D.M. Newns, P.C. Pattnaik, H. de Raedt and H.G. Matuttis
for useful discussions.
Part of this work was supported by the Deutsche Forschungsgemeinschaft (DFG).
We are grateful for the Leibnitz Rechenzentrum M\"unchen (LRZ) for providing
a generous amount of CPU-time on their IBM SP 2.

\begin{table}
\begin{tabular}{cccccc} 
    U    &  $E_{ED}/N $ & $E_{SD}/N $ & $E_{single}/N $ & $E_{slice}/N$ & 
$E_{cluster}$ \\ \hline
   2.0   & -1.336059 & -1.3360 & $-1.337\pm 0.001$ & $-1.335\pm 0.001$ &$ -1.332 \pm 0.003 $\\
   4.0   & -1.223808 & -1.2237 & $-1.224\pm 0.002$ &  $-1.224 \pm 0.002$ & $-1.22\pm 0.01$\\
   5.0   & -1.182000 & -1.1819 & $-1.180\pm 0.003$ &  $-1.184 \pm 0.003$ & $-1.19\pm 0.02$   \\
\end{tabular}
\caption{Comparison of the ground state energies per lattice site
for a Hubbard model of system size
$N=4\times 4$, number of electrons $N_e=5+5$, diagonal hopping-matrix element $t_p=0.0$
for various interactions $U$. In column 1 we show 
the interaction $U$, 
in column 2 we present the results for the exact diagonalization  ($E_{ED}/N $),
in column 3 we show the stochastic diagonalization results ($E_{SD}/N$), 
in column 4-6 we present the
results obtained with the PQMC-Algorithms for different elementary moves.
Details of these moves are given in the text.}
\label{tab1}
\end{table}

\begin{table}
\begin{tabular}{cccccc} 
    U    &  $E_{ED}/N $ & $E_{SD}/N$ & $E_{single}/N $ & $E_{slice}/N$ & $E_{cluster}/N$\\ \hline
   2.0   & -1.230034  & -1.2300  & $-1.2301\pm 0.0003$ & $-1.2306\pm 0.0008$ & $-1.229 \pm 0.001$ \\
   4.0   & -1.126160  & -1.1261  & $-1.125\pm  0.003$ & $ -1.125\pm 0.002$ &  $-1.13\pm 0.02$ \\
   5.0   & -1.088907  &-1.0887   & $-1.089\pm  0.004$ & $ -1.087\pm 0.004$ &  $-1.10\pm 0.02$\\
   6.0   & -1.058717  &-1.0581  &  $-1.061\pm  0.005$ & $ -1.058\pm 0.007$ &  $-1.06\pm 0.02$\\
\end{tabular}
\caption{Comparison of the ground state energies per lattice site 
for a Hubbard model of system size
$N=4\times 4$, number of electrons $N_e=5+5$, diagonal hopping-matrix 
element $t_p=-0.22$ for various interaction $U$. In column 1 we show 
the interactions $U$, 
in column 2 we present the results for the exact diagonalization  ($E_{ED}/N $),
in column 3 we show the stochastic diagonalization result 
($E_{SD}/N$), in column 4-6 we present the
results obtained with the PQMC-Algorithms for different elementary moves.
Details of these moves are given in the text.}
\label{tab2}
\end{table}

\begin{table}
\begin{tabular}{ccccc} 
           & $U=2$ &   $U=4$ & $U=5$ & $U=6$ \\  \hline
$t_p=-0.22$& 1.0   & 0.92    & 0.69  & 0.42  \\
$t_p=0.0$  & 1.0   & 0.99    & 0.95  & 0.85  \\
\end{tabular}
\caption{Average sign $\langle {\rm sign} \rangle$ for $4 \times 4$ Hubbard cluster with
$N_e=5+5$, $\langle {\rm sign} \rangle$ is in the given precision the same in all three
implementations of the PQMC algorithm.}
\label{tab3}
\end{table}

\begin{table}
\begin{tabular}{crrr} 
 U    & cluster &   single & slice \\  \hline
  2.0 & 790\,000   &  49\,000 & 23\,400 \\
  4.0 & 890\,000   &  79\,000  & 50\,000 \\
  5.0 & 4\,490\,000   &  199\,000  & 65\,200 \\
  6.0 & 5\,390\,000   &  599\,000  & 72\,900  \\ 
\end{tabular}
\caption{
\label{tabMCSpos}
Monte Carlo Sweeps MCS for cluster and single spin flip
algorithm for the a $4 \times 4$ Hubbard model with
$N_e=5+5$, $t_p=-.22$ and interaction $U$ of the runs in figure 
1, 2, 4 and 5 }
\end{table}

\begin{table}
\begin{tabular}{cccccc}
    U    &  $E_{ED}/N $ & $E_{SD}/N$ & $E_{single}/N $ &  $E_{cluster}/N$\\ \hline
  -2.0   & -1.731689  & -1.7316  &  $-1.731\pm 0.003$ & $-1.732 \pm 0.006$ \\
  -4.0   &  -2.045849 & -2.0453  & $ -2.045\pm 0.002$ &  $ -2.048\pm 0.01$ \\
  -6.0   & -2.458782  &-2.4568   & $ -2.460\pm 0.004$ &  $ -2.47\pm 0.01$\\
  -8.0   & -2.956890  &-2.9545   & $ -2.952\pm 0.01$ &  $ -2.969\pm 0.05$\\
\end{tabular}
\caption{Comparison of the ground state energies per lattice site 
for an attractive Hubbard model of system size
$N=4\times 4$, number of electrons $N_e=5+5$, diagonal hopping
matrix element $t_p=0$
for various interactions $U$. In column 1 we show
the interaction $U$, in column 2 we present the results for the exact diagonalization  ($E_{ED}/N $),
in column 3 we show the stochastic diagonalization result
($E_{SD}/N$), in column 4-5 we present the
results obtained with the PQMC-Algorithms for different elementary moves.
Details of these moves are given in the text.}
\label{tab5}
\end{table}

\begin{table}
\begin{tabular}{crrcr} 
$U$ & $t_p$ & $N_{SD}$ &  $N_1^{(n,m)}$ & $T_A$ \\ \hline
$6.0$ & $-.22$ & 307\,324 & $0.0491$ & $0.381 \cdot 10^{-7}$ \\  
$5.0$ & $-.22$ & 285\,872 & $0.0426$ & $0.381 \cdot 10^{-7}$  \\
$4.0$ & $-.22$ & 228\,172 & $0.0334$ & $0.381 \cdot 10^{-7}$ \\ 
$2.0$ & $-.22$ & 141\,742 & $0.0062$ & $0.954 \cdot 10^{-8}$ \\
$1.0$ & $-.22$ &  60\,638 & $0.0020$ & $0.238 \cdot 10^{-8}$ \\
$-1.0$ & $0.0$ & 20\,974 & $0.0040$ & $0.954 \cdot 10^{-8}$  \\
$-2.0$ & $0.0$ & 28\,034 &  $0.0094$  & $0.298 \cdot 10^{-8}$ \\
$-2.0$ & $0.0$ & 115\,304 &  $0.0140$  & $0.191 \cdot 10^{-7}$ \\
$-4.0$ & $0.0$ & 179\,755 & $0.1117$ &  $0.153 \cdot 10^{-6}$ \\
$-6.0$ & $0.0$ & 253\,275 & $0.1297$ & $0.153 \cdot 10^{-6}$  \\
$-8.0$ & $0.0$ & 263\,862  & $0.1514$ & $0.153 \cdot 10^{-6}$  \\
\end{tabular}
\caption{Number of states $N_{SD}$ used for the stochastic diagonalization 
method in a $4\times 4$ system with $N_e=5+5$ electrons and $t_p=0$ 
for different interaction $U$. $N_1^{(n,m)}$ is the norm of the 
first column in the transformed matrix $H^{(n)}$, and $T_A$ is the 
threshold for the acceptance of a new basis states to the basis of 
important states.
\label{tabstates}
}
\end{table}

\begin{table}
\begin{tabular}{ccccc} 
$nstab$ & MCS &   $E_{cluster}$ &  \\  \hline
1  &  1745000  &  $-2.45282 \pm  0.00607 $ \\
2  &  1745000  &  $-2.45282 \pm  0.00607 $ \\
4  &  1745000  &  $-2.45282 \pm  0.00607 $ \\
8  &  1745000  &  $-2.45282 \pm  0.00607 $ \\
16  &  1745000  &  $-2.45282 \pm  0.00607 $ \\
\end{tabular}
\caption{Groundstate energy for $4 \times 4$ Hubbard cluster with
$N_e=5+5$, $t_p=0.0$ and $U=-6$ for different numbers of stabilizations
$nstab$.}
\label{tab10}
\end{table}

\begin{table}
\begin{tabular}{crr} 
 U    & cluster &   single  \\  \hline
 -2.0 & 39\,000   &  4\,000  \\
 -4.0 & 189\,000   &  9\,000   \\
 -6.0 & 2\,909\,000   &  109\,000   \\
 -8.0 & 7\,750\,000   &  149\,000    \\ 
\end{tabular}
\caption{
\label{tabMCS}
Monte Carlo Sweeps MCS for the cluster and the single spin flip
algorithm for a $4 \times 4$ Hubbard model with
$N_e=5+5$, $t_p=0$. The interactions $U$ are the same of the runs 
in figure 4 and 5.}
\end{table}

\begin{table}
\begin{tabular}{cccccc}
    U    &  $E_{ED}/N $ & $E_{SD}/N$ & $E_{single}/N $ &  $E_{cluster}/N$\\ \hline
  -2.0   & -1.731689  & -1.7314  &  $-1.731\pm 0.003$ & $-1.732 \pm 0.006$ \\
  -4.0   & -2.045849  & -2.0419  & $ -2.035\pm 0.003$ &  $ -2.078\pm 0.$ \\
  -6.0   & -2.458782  & -2.4451  & $ -2.457\pm 0.002$ &  $ -2.439\pm 0.$\\
  -8.0   & -2.956890  & -2.9432  & $ -2.952\pm 0.001$ &  $ -2.977\pm 0.$\\
\end{tabular}
\caption{Comparison of the ground state energies per lattice site 
for an attractive Hubbard model of system size
$N=4\times 4$, number of electrons $N_e=5+5$, diagonal hopping
matrix element $t_p=0$ for various interactions $U$. 
Here we compare unconverged runs.
In column 1 we show the interaction $U$, in column 2 we present the 
results for the exact diagonalization  ($E_{ED}/N $),
in column 3 we show the stochastic diagonalization result
($E_{SD}/N$), in column 4-5 we present the
results obtained with the PQMC-Algorithms for different elementary moves.
Details of these moves are given in the text.}
\label{tabbadenergy}
\end{table}

\begin{table}
\begin{tabular}{crrr} 
 U    & state in SD & cluster &   single  \\  \hline
 -2.0 & 12\,122 & 39\,000  &  4\,000  \\            
 -4.0 & 23\,510 & 50\,000  &  5\,000   \\           
 -6.0 & 32\,587 & 1\,000\,000   &  49\,000   \\     
 -8.0 & 46\,583 & 1\,500\,000   &  109\,000    \\   
\end{tabular}
\caption{
\label{tabMCSunconverg}
Number of basis states in the SD and Monte Carlo Sweeps MCS for the 
cluster and the single spin flip
algorithm for a $4 \times 4$ Hubbard model with
$N_e=5+5$, $t_p=0$ and attractive interaction $U$ of the unconverged 
runs of figure 6.}
\end{table}

\end{document}